# NEMESIS RECONSIDERED

Adrian L. Melott[1] and Richard K. Bambach[2]

1. Department of Physics and Astronomy, University of Kansas, Lawrence, KS 66045 USA

2. Department of Paleobiology, National Museum of Natural History,

Smithsonian Institution, PO Box 37012, MRC 121, Washington, DC 20013-7012 USA

ABSTRACT

The hypothesis of a companion object (Nemesis) orbiting the Sun was motivated by the claim of a terrestrial extinction periodicity, thought to be mediated by comet showers. The orbit of a distant companion to the Sun is expected to be perturbed by the Galactic tidal field and encounters with passing stars, which will induce variation in the period. We examine the evidence for the previously proposed periodicity, using two modern, greatly improved paleontological datasets of fossil biodiversity. We find that there is a narrow peak at 27 My in the cross-spectrum of extinction intensity time series between these independent datasets. This periodicity extends over a time period nearly twice that for which it was originally noted. An excess of extinction events are associated with this periodicity at 99% confidence. In this sense we confirm the originally noted feature in the time series for extinction. However, we find that it displays extremely regular timing for about 0.5 Gy. The regularity of the timing compared with earlier calculations of orbital perturbation would seem to exclude the Nemesis hypothesis as a causal factor.

1. INTRODUCTION

Raup and Sepkoski (1984), hereafter RS, published a study of marine fossil families, describing in it a 26 My periodicity in local maxima of extinction; Sepkoski (1989) confirmed and refined it to 26.2 My with a study of genera; Lieberman & Melott (2007) found hints of it with spectral analysis. This result cannot be said to have been settled by consensus among paleontologists. We will not review this controversy, but will add new information to the examination of evidence for it. Our primary purpose is to ask whether such evidence as exists is consistent with the idea of a dark companion to the Sun, often called Nemesis, which causes the periodic pattern of extinction. This is closely related to goals of the WISE, LSST, and Pan-Starrs all-sky surveys. A slight difference in period with the 1984 RS result is expected, since the modern geological timescale (Gradstein et al. 2004) is expanded about 3% over the same time period RS used. Consequently we test the hypothesis of a 27 My periodicity. It should be anchored at the K-T extinction event, which specifies the phase. This is a new, independent test with a specified phase and period of the RS hypothesis based on expanded data and a new geological timescale.

Nemesis was proposed by Davis, Hut, and Muller (1984) and Whitmire and Jackson (1984) based on the idea that a companion star in a wide orbit perturbs the Oort cloud every 26 My, causing comets to enter the inner solar system, some of which collide with the Earth potentially causing mass extinctions. There is considerable parameter space in which such an object could exist (e.g. Iorio 2009). One of the serious questions is whether such a distant object could have a stable orbit. There is extensive discussion of this question in Muller (2002). Basically, the orbit perturbation is expected to change the period a few My each orbit. At some point perhaps 1.5 Gyr in the future, it becomes unbound.

We show here that the fossil record is inconsistent with perturbations expected in the orbit of a dark Solar companion with the requisite orbital period.

2. EXPECTED ORBITAL PERTURBATIONS

There was considerable discussion of the orbital stability of a Nemesis candidate around the time that the idea was proposed. Two sources of perturbation have been considered: perturbation from the Galactic tidal gravitational field and perturbations by passing stars. Hut (1984) was specific that irregularity of the period of revolution of such an object over the past 250 My should be about 20% due to perturbation from the Galaxy tidal gravitational field and by passing stars, and sharp peaks should not be expected in Fourier analysis. Torbett & Smoluchowski (1984) reached the same conclusion, but with a somewhat larger estimate of the fluctuations from the Galactic tide alone, dependent on the inclination of the Nemesis orbit with respect to the Galactic disk. Hills (1984) estimated a period change of 4% per Nemesis orbital period from the effects of passing stars. Using a $t^{1/2}$ amplitude scaling expected from a random walk, the orbital period should drift by 15 to 30% over the last 500 My. This change in the period will broaden or split any spectral peak in a time series frequency spectrum, so Nemesis as an extinction driver is inconsistent with a sharp peak (e.g. Muller and MacDonald 2002).

3. DATA AND METHODOLOGY

There is a large literature on the analysis of the time evolution of fossil biodiversity and of extinctions. We have for some time conducted a study (Melott and Bambach 2010) of the evidence for a strong 62 My periodicity in fossil biodiversity This was discovered by Rohde & Muller (2005), and is apparently not related to the proposed 26.2 My signal. Here we use modern data and a cross-comparison of two data sets to extend our techniques to examine periodicity in extinction.

There is an extensive discussion of our methodology in Melott and Bambach (2010), which we will summarize here, as applied to the extinction data. We use two compendia: (1) The data set of Sepkoski (2002), with the cuts made to determine "Well-resolved fauna" by Rohde & Muller (2005). (2) The Paleobiology Database data as published in Alroy et al. (2008); see also Alroy (2008). Both are treated with the most current geological timescale, that of Gradstein et al. (2004). Both are compendia of marine fauna—the standard paleontological dataset for diversity history studies, due to the quality of the record.

The Sepkoski data is essentially a compendium of all the data available to the author, so that it emphasizes quantity of data. The Paleobiology Database was selectively subsampled so as to maintain a consistent sampling rate over time and over geographical location; consequently the numbers available in any given realization are considerably smaller. The mean time interval length in the Sepkoski data is about 3.6 My, and about 11 My in the Paleobiology Database set. We examine the extinction of genera in these data, defined as the time of the last appearance of any member of the genus in the given fossil record. The overall usable record length is approximately 500 My, the length of time since substantial hard body parts evolved, capable of leaving a rich fossil record.

The difference in style between these data sets seems similar to the dichotomy between the largest available extragalactic redshift surveys and the smaller, magnitude-limited ones emerging c. 1980. There were very different kinds of conclusions about superclusters, which eventually converged as the magnitude-limited surveys became larger.

Extinction intensity, i.e. the number of genera going extinct divided by the number in existence at the given stage is tabulated in the two data sets. We assign the extinction to the end date of the period, as the data better support this than a continuous rate (Foote 2005); also this is appropriate to the investigation of a pulsed causality model as considered here. In order to use the function as available in both sets, when we compute cross-spectra (1) They are truncated to the period of time between 23 and 467 My ago. At one end divergences due to small numbers of genera and at the other the process of sample-standardization in the Paleobiology Database data preclude comparative analysis extending over the full range. (2) As needed for Fourier analysis (Muller and MacDonald 2002), both data sets are detrended as shown elsewhere (Rohde and Muller 2005; Cornette 2007; Melott and Bambach 2010). At frequencies considered here the results are not sensitive to the choice of detrending function (a cubic), within polynomials and exponential functions. Each series is divided by its own standard deviation to put them on an equal footing.

A major strategy for suppressing random error lies in constructing a cross-spectrum of the two data sets. Results that appear identical in multiple independent data sets are a strong argument against their being random fluctuations. We are interested in whether the same frequencies are strong in both data sets, with the same phase, for the two detrended series. The cross-spectrum is most easily described using the complex exponential representation of Fourier Series. It is a generalization of the power spectrum of a single time series, which is essentially $A_i^*A_i$, where * denotes complex conjugation and $A_i$ denotes elements of a series of complex Fourier coefficients as a function of frequency. The power spectrum is therefore real-valued. The cross-spectrum involves the Fourier coefficients of two different series: $B_i{}^*C_i$, and is complex. The amplitude of this complex number is a measure of the extent to which a given frequency is present in both series; its phase is a measure of the extent to which the components of the two series are in phase. To the extent that these signals are anchored in reality, we expect phase agreement between components from the two data sets. If they are in perfect phase agreement, the cross-spectral coefficient is real and positive. Thus the test for a positive peak in the real part of the cross-spectrum is a stringent test that requires not only that both data sets have the same frequency at strength but also that the signals have some component in phase—peaks and troughs coincide in time. All data were linearly interpolated between adjacent data points to give results at 1 My intervals. This effective window function is has minimal effects at frequencies considered here (Melott and Bambach 2010). ALM constructed the cross-spectra and checked them against the independent power spectra of the two samples, as well as the non-interpolated data with Lomb-Scargle (e.g. Scargle 1982). We found that a signal at 27 Myr appeared in both data sets separately, but the suppression of random error in the combined set as well as a comparison with mass extinction timings (Part 4.2), clarified things greatly.

4. RESULTS

4.1 Cross-spectrum

In Figure 1 we show on the y-axis the real part of the cross-spectrum of extinction intensity as described above. The x-axis is the frequency in $My^{-1}$. There are three prominent peaks. There is a peak to the left at a period of 99.9 My, which we do not address, a larger one at 62 My, which has been extensively discussed as a major periodicity in total fossil biodiversity elsewhere (Rohde & Muller 2005; Melott & Bambach 2010 and references therein), and a third, smaller but quite sharp peak at 26.8 My. The first two are interesting, but not relevant to the Nemesis question. The secondary sharp peak corresponding to a period of 26.8 My is consistent with the prediction of a 27 My feature based on RS. The peak is very narrow, only positive for periods between 25.7 and 28.2 My, or less than 10% full width. The width of the power spectral feature in each data set by itself is the same, basically ±1 My, and their peaks are at 26.97 and 27.13 My. The spectral amplitude at 26.8 My PBDB alone is 0.033, and in the Sepkoski data alone is 0.039 (after normalization). The amplitude difference between the two can be understand as a sampling effect (Melott and Bambach 2010), since the mean interval length in PBDB is about 11 My. We also found a peak at 27 My using only Sepkoski data older than 250 My, which was a time period not included in the original RS study.

At the spectral peak, the phase difference between the two signals at the indicated spectral peak corresponds to 3 My. The coherency γ = 0.77 agrees well with the implied phase difference (Muller and MacDonald 2002). The significance of this coherency can be computed from the formula of Goodman (1957); see also Thompson (1979). This is based on the normalized amplitude of the peak, compared with the peaks in the individual power spectra. We use n=35 degrees of freedom, the number of independent values over the included time range in PBDB, the lower time resolution sample. The coherency value found here has a p-value 2 X $10^{-14}$. This gives the confidence that a random fluctuation has not put the two series in synch at this spectral peak. So we can be confident that they are both seeing the same periodicity. What about its amplitude?

We have plotted the function showing negative real values as well. Negative real values indicate oppositely directed trends in the two datasets which are a good indicator of random error. The maximum amplitude of the negative fluctuations is about 0.005, which is one way to estimate background fluctuations. As can be seen in Figure 1, the peak in question is about five times greater in amplitude than the negative fluctuations. Systematic error is more problematic, but the Paleobiology Database set was constructed specifically with the purpose of suppressing systematic error. Both data sets individually show a peak at 27 Myr, but their combination will suppress random error. This narrow peak is inconsistent with either a broadened peak, expected from Nemesis with continuous perturbations, or a split peak if there were discrete changes in period (Muller and MacDonald 2002).

4.2 Relation to mass extinction events

In Figure 2 we show how the periodicity is visible in the extinction intensity of the Sepkoski data (the higher time resolution set) as a function of time. Overlaid on it are vertical lines corresponding to a period of 27 My. It may be noted that a great many peaks of extinction intensity lie at or near the 27 My period. This correspondence may be treated more precisely. Bambach (2006) reviewed extinction and identified 19 "mass extinctions" with selection criteria of magnitude and rapidity. These are circled in Fig. 2. This is a pre-defined set, so we can ask about their timing with respect to the 27 My periodicity. We anchored our series at the K-T extinction event 65 My ago, the most major "recent" mass extinction, the same as RS. Although we did not use this procedure to select our phase angle or period, after the fact we verified that this anchor for the series maximizes the number of "hits", i.e. coinciding with the maximum number of the preselected mass extinction events. We also verified that it maximizes hits for the Paleozoic period (≥ 251 My ago), which RS did not include. The dashed lines in Figure 2 begin at 524 My ago and are spaced 27 My apart with the last 11 My ago.

We now use this to test the null hypothesis that there is no excess of extinctions clustered at a 27 My interval with respect to random. Note that finding an excess at 27 My does not imply that all extinctions should happen at these intervals, nor that this is the strongest periodic signal in the data.

As mass extinction events are taken to be impulsive, or nearly so, the following statistics are more appropriate for determining significance of timing than are power spectra. We find that 10 of the 19 mass extinction events lie within 3 My either side of the 27 My timing lines. Since the position of the line series is determined by one event, we lose one degree of freedom. Therefore, this is equivalent to hitting 9 out of 18 with a single-event random probability of 6/27 or 0.222... Application of the binomial theorem gives the probability that this can be a random result at about 0.0088, giving us slightly better than 99% confidence in rejecting the hypothesis that there is no association of the mass extinctions with the 27 My periodicity. If the tolerance is shifted to 2 My, the number of hits is decreased but the p-value stays around 0.01.

An alternate way to run the test is to consider all local maxima (many of which are peaks with low amplitude) in Fig. 2, i.e. all points that have extinction intensities greater than those on either side. This uses the same phase angle as the previous test. Losing one degree of freedom for this, there 16 hits out of 43, which is still in excess of random, and gives a p-value of 0.019.

The period 27 My is significant on the basis of two different tests, using two sets of data, examining either all extinction local maxima or those previously classified as "mass extinctions", and clearly shows no detectable drift in either period or phase. We have detected something equivalent to the RS extinction periodicity, but extending over nearly twice the duration. The p-value is about 1%, exceeding their original claim of 5%. We have doubled the time interval over which the extinction intensity periodicity is detected, fixed the revised period close to 27 My, and increased the confidence level from 95% to 99%.

## 5. DISCUSSION AND IMPLICATIONS FOR NEMESIS

The peak we have found is measured over ~500 My (possible with modern paleontological data) and appears with a confidence level of p=0.01 with two different statistical tests based on extinction intensity, and p=0.02 on one based on "lesser peaks" of this function. It shows less than 10% variation in period by a spectral test, and less than 2% by an extinction timing test, over the entire time period. In fact our cross-spectral peak has the narrowest bandwidth possible consistent with the level of probable random error in the fossil record. Fossil data, which motivated the idea of Nemesis, now militate against it and suggest another mechanism is needed to explain extinction periodicity. An attempt to associate the periodicity with passage through the Galactic mid-plane (e.g. Clube and Napier 1984) has its own set of problems: 54 My is rather too short for most estimates of the period of the Sun normal to the plane, and our passage within the last My or so of the mid-plane (e.g. Gies and Helsel 2005) is inconsistent with the phase of the 27 My signal we have detected, with its recent maximum at 11 My ago.

## ACKNOWLEDGMENTS

We thank J. Alroy for the Paleobiology Database extinction data, and R. Muller for his sort of the Sepkoski data and useful discussions about Nemesis. We are grateful to the American Astronomical Society for the sponsorship of the 2007 Honolulu multidisciplinary Splinter Meeting at which discussions leading to this project took place. J. Cornette and an anonymous referee provided useful input. Research support at the University of Kansas was provided by the NASA Program Astrobiology: Exobiology and Evolutionary Biology under grant number NNX09AM85G.6

FIGURE CAPTIONS

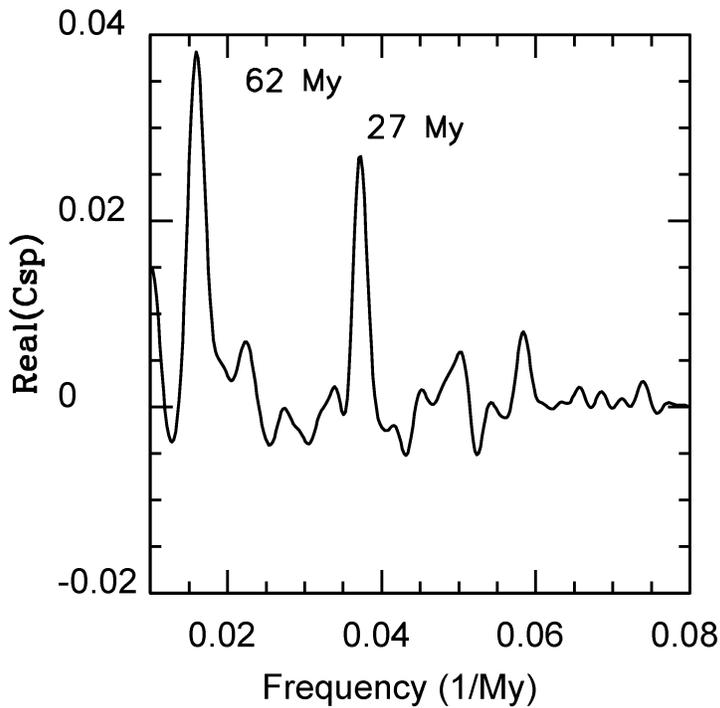

Figure 1. The real part of the cross spectrum of fluctuations of extinction intensity against frequency in My$^{-1}$ for the two data sets used here. The large peak on the left corresponds to a period of 62 My in total biodiversity, studied elsewhere. The narrow peak on the right corresponds to a period of 27 My in phase with the original Raup and Sepkoski (1984) extinction periodicity which led to the suggestion of Nemesis. Negative values are indicative of random error, and positive values of phase agreement between the independent data sets.

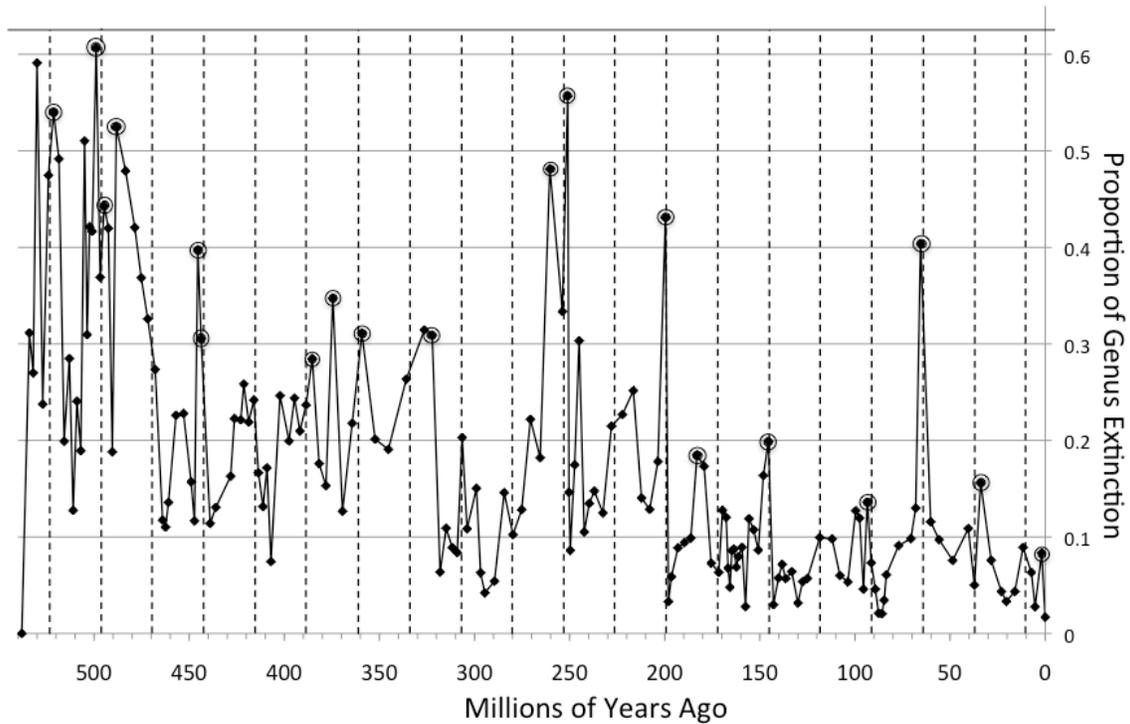

Figure 2. A plot of extinction intensity (genera going extinct divided by total extant genera) at the substage level from the Sepkoski data. Extinction dates are taken as end dates of intervals as described in the text. Events meeting the criteria for "mass extinctions" in the review of Bambach (2006) are circled. It is visually apparent that an large proportion of the circled points lie near the vertical lines which mark off a 27 My interval. In the text we show a probability of p≤0.01 that this can be a random alignment.